\documentclass[12pt]{article}
\usepackage{amsmath,amssymb}
\usepackage{ifpdf}
\makeatletter

    \@addtoreset{equation}{section}
\makeatother
\ifpdf
  \usepackage[pdftex]{graphicx}
\else
  \usepackage[dvipdfm]{graphicx}
\fi

\newcommand{\One}{{\hbox{{\rm 1{\hbox to 1.5pt{\hss\rm1}}}}}}

\newcommand\be{\begin{equation}}
\newcommand\ee{\end{equation}}
\newcommand\bea{\begin{eqnarray}}
\newcommand\eea{\end{eqnarray}}
\newcommand\ba{\(\begin{array}}
\newcommand\ea{\end{array})\ }
\newcommand\nn{\nonumber}

\begin{document}
\thispagestyle{empty}
\vspace*{2cm}
\begin{center}
 {\LARGE {Irregular conformal block and its matrix model}}
\vskip1.5cm
{\large 
 Chaiho Rim\footnote{email: rimpine@sogang.ac.kr}
}
\vskip.5cm
{\it Department of Physics and Center for Quantum Spacetime (CQUeST)
\\
Sogang University, Seoul 121-742, Korea}
\end{center}
%
%
\begin{abstract}
Irregular conformal block is a new tool to study Argyres-Douglas theory,
whose  irregular vector is represented as a simultaneous 
eigenstate of a set of positive Virasoro generators. 
One way to find the irregular conformal block 
is to use the partition function of the $\beta$-ensemble 
of  hermitian matrix model. 
So far  the method is limited to the case of irregular
singularity of even degree. 
In this letter, we present a new matrix model  for the case of odd degree 
and calculate its partition function. 
The model is different from the previous one 
in that its potential has additional factor of square root of matrix.  

\end{abstract}
%
%
\tableofcontents
%

\section{Introduction}

Irregular conformal block (ICB) 
is obtained from the conformal block by taking the colliding limit
which puts some of primary fields at the same point 
maintaining its limit finite  \cite{GT:2012}. 
ICB generalizes the AGT relation \cite {AGT:2009} 
and is considered to  reproduce \cite{GT:2012, Bonelli:2011aa}  
the Argyres-Douglas (AD) theory  \cite{AD:1995}. 

ICB is first obtained as the eigenstate of  Virasoro generator 
$L_1$ or  $(L_1, L_2)$ in \cite{Gaiotto:0908}. 
The eigenstate is not annihilated by $L_k$ with $k>0$ but  
is an eigenstate.  This coherent state is called irregular vector. 
This idea is generalized so that 
the irregular vector is the simultaneous eigenstate of 
$L_k$'s  ($0< n\le k  \le 2n $) 
and  is constructed in terms of the free field representation 
in the presence of background charge in \cite{GT:2012}. 
In fact, this ICB  is described by the Seiberg-Witten curve 
$x^2 = \phi_2$ where $\phi_2$ is a quadratic differential 
with  a pole of degree $2n+2$,
which is associated to a punctured Riemann sphere 
and is classified as $D_{2n} $ of AD theory
\cite{Cecotti:2011rv}. 

On the other hand, it is  noted in \cite{TR:2012} that 
ICB is conveniently obtained from the hermitian matrix model 
whose potential consists of inverse powers of  matrix 
and logarithmic one.  
Its partition function is identified 
with the inner product of an  irregular vector at the origin and a
regular one at infinity. 
Because this matrix model can be obtained 
as the appropriate limit of the Penner-type 
matrix model \cite{Dijkgraaf:2009pc, Eguchi:2009gf, Nishinaka:2011aa},
this matrix model approach shares the same idea 
as that of the colliding limit of the regular conformal block. 

The matrix model is, however, so far limited to the case of even degree.  
The presence of the pole of degree $2n+1$ 
requires the irregular state annihilated by $L_{2n}$
but  simultaneous eigenstates of  $L_k$ with $0< n\le k  \le 2n-1 $. 
This ICB may be constructed in terms of the twisted free field modes 
which  introduce a new square root branch-cut at the origin
as shown in  \cite{GT:2012}.  

In this letter, we present a new matrix model with square root  potential
and calculate the corresponding partition function.  
In section \ref{partitionfunction}, we define the matrix model 
and obtain the loop equation.  The square root potential not only introduces
the branch-cut at the origin but also changes the details of the loop equation. 
In section \ref{solution}, we provide the solution to the simplest non-trivial case 
namely, the case $n=2$ (or $D_3$) explicitly.  
It is straightforward to generalize to the case with $n>2$.
Section \ref{discussion} is for discussion and 
a detailed derivation of the loop equation is provided 
in appendix \ref{f(z)}. 

\section{Partition function and loop equation} 
 \label{partitionfunction}

We consider the  partition function of a hermitian matrix
\be
Z_M= \int \left[ \prod_{I=1}^N dz_I \right]
 \Delta(\lambda_I)^{2\beta} e^{\frac{\sqrt{\beta}}g \sum_I V(\lambda_I)}
\ee
where $ \Delta(\lambda_I)$ is the Vandermonde determinant.  
$\lambda_I$ is the eigenvalue of the matrix and $N$ is the size of the matrix.   
The potential  has the form  
\begin{align} 
 V(z)= \alpha \log z  - \sum_{s \in S}  \frac{  c_{s}}{s  z^{s} } 
\label{Vz}
\end{align}  
where $S=\{1/2, 3/2, \cdots,  n-1/2 \}$ is a finite set of  $n$ half-odd integers.
It is convenient to denote the potential as $V(z) = V_e(z) + V_o(z)$ 
where $ V_e(z)= \alpha \log z $ and $ V_o(z)= - \sum_{s \in S}  \frac{  c_{s}}{s  z^{s} }$ 
so that $ V_o(z)$ has a cut $\Gamma$ along the negative real axis.  
We assume that there is an appropriate domain of the parameter space  $\{  \alpha, c_S\}$
so that the partition function is well defined 
when the integration range lies along the positive real axis.

The saddle point equation holds on the positive real axis  
\be
2 w(\lambda_I) + V'(\lambda_I) =0\,,~~~~
w(\lambda_I) = g \sqrt{\beta} \sum_{J \ne I} 
\left( \frac1{\lambda_I - \lambda_J} \right) \,. 
\label{saddle} 
\ee 
A nice way to find the functional relation is to introduce the resolvent  
\cite{resolvent},
$ W(z) =g  \sqrt{\beta} 
\left\langle \sum_{J}\frac{1 }{z-\lambda_J } \right\rangle$. 
Here $\langle \cdots \rangle$ stands for the expectation value
with the given potential $V(z)$.
The saddle point equation \eqref{saddle} 
shows that  the resolvent is discontinuous 
along  a certain integration range $\Lambda$ of the partition function
\be
W(\xi+ i 0) + W(\xi -i 0) +  V'(\xi) =0 ~~~{\rm when}~\xi \in \Lambda\,.
\label{lambda-discontinuity}
\ee 
This  discontinuity is  encoded in terms of $ G (z) \equiv ~ 2 W(z)+  V'(z) $
so that $ G(\xi- i\epsilon) +  G(\xi+ i \epsilon)  =0$ is automatically satisfied 
on the square-root branch cut. 
Furthermore, $ G (z)$ is not continuous on the branch cut $\Gamma$
and consists of the monodromy even and odd parts.  

The resolvent satisfies the loop equation of the form
\be
f(z) =4 W(z)^2  +4 V'(z)W(z) + 2 \hbar QW'(z)    - \hbar^2  W(z,z) \,.
\label{loop-eq}
\ee
where we switch the notation from $g$ and $\beta$ to  
$\sqrt{\beta}= -ib$, $\hbar= -2ig$ and$Q=b + 1/b$.  
$  W(z_1,z_2) = \beta   \left\langle  \sum_{I_1, I_2}  \frac{1}{z_1-\lambda_{I_1}} \frac{1}{z_2-\lambda_{I_2}} \right\rangle_{\!\!\rm conn} $  is a two-point  (connected) resolvent  
and $f(z)$ is the quantum correction defined as 
$  f(z) = 2\hbar b  \sum_{I=1}^N\left\langle \frac{V'(z)-V'(\lambda_I)}{z-\lambda_I}\right\rangle$.
As  shown in appendix \ref{f(z)}, its explicit form is 
\begin{align}
 f(z)&= \sum_{k=0}^{n-1}  \frac  {d_k } {z^{2+k} }  + 4  \frac{V_o'(z)}{\sqrt{z}}  ~ \tau(z) 
\label{f(z)-form}
\end{align}
where  $d_k =  v_k  (- \hbar^2 \log Z_M ) $ 
and $v_k = \sum_{s \in S}  s c_{s +k}   \frac{\partial} {\partial c_s } $. 
Here we use the definition $c_s =0$ if $s$ does not belong to the set $ S$.  
$ \tau(z)  =  \frac{\hbar b}2 \sum_I   \left \langle  \frac 1{\sqrt{z} +\sqrt{\lambda_I}} \right \rangle$
is a new analytic function, continuous on  $ \Lambda$  but discontinuous on $\Gamma$.

\section{Solution to the lowest order in $\hbar$}
\label{solution} 

The coefficient $d_k$ in \eqref{f(z)-form} is the key element to find the partition function.
Once $d_k$'s are known, one can find the partition function using the differential equation
with respect to the parameters of the potential. 
Our strategy is to find $d_k$'s from the loop equation order by oder in $\hbar$.
To make the expansion feasible, we assume 
that  $ V'(z)  $ is the order of $\hbar$
so that the  expansion is  equivalent to  the large $N$ expansion  
($\hbar \propto 1/N$).
This is achieved if  $\alpha $ and $c_S$ are proportional to $\hbar$. 

To the lowest order in $\hbar$ (planar limit), 
$G(z)$ satisfies the loop equation \eqref{loop-eq}
\be
 G(z)^2 -   2V_o'(z) H(z)  = \varphi_2(z)
\label{loop-G}
\ee
where $H(z)  = 2 {\tau(z)}/  {\sqrt{z} } + V_e'(z) $ and 
$  \varphi_2(z) = \sum_{k=0}^{n-1}  \frac  {d_k } {z^{2+k} }    + V_e'(z)^2 + V_o'(z)^2$
is the expectation value of the energy-momentum tensor 
which has the pole of  odd degree $2n+1$. 
At large $z$, the dominant term of \eqref{loop-G} is the order of $1/z^2$ 
and its coefficient provides an identity 
\be
 (\hbar b N + \alpha)^2 = d_0 + \alpha^2\,.
\label{d_0}
\ee

For the simplest case, $n=1$, 
$ d_0 = \frac12 c_{1/2} \frac{\partial}{\partial c_{1/2}} (- \hbar^2 \log Z_M) $
determines the partition function,
$ Z_M =\zeta~   ( c_{1/2}) ^{-2\left(  (b N  )^2 + 2 b N \alpha/ \hbar \right) }$.
$\zeta$ is a $ c_{1/2}$-independent constant and can be absorbed to the 
definition of the partition function. 
Furthermore, one has 
$G(z) = \frac{\hbar b N + \alpha}z + \frac{c_1/2}{z^{3/2}}$ 
and 
$ H(z) = \frac{\hbar b N + \alpha}z $ 
which shows no  no branch cut $\Lambda$. 
In fact, $\Lambda$ reduces to the origin $z=0$ 
as seen from the  $n=2$ case below (if $c_{3/2}$ is turned off). 

When  $n\ge 2$, the solution is more involved.  
It is natural to assume there are $n-1$ cuts $\Lambda_k$'s
each of which lies between $a_k$ and $b_k$ ($0< a_k<b_k$)
on the positive real axis. 
In addition, according to the monodromy at $z=0$,
one may put $G(z) = G_e(z) + G_o(z)$;
\be
G_e(z) = \frac{ g_e(z)}{z^n}  {\prod_k^{n-1}\sqrt  {(z-a_k)(z-b_k)}}  \,,~~~
G_o(z) =  \frac{ g_o(z)}{ z^{n+1/2}}  \prod_k^{n-1} \sqrt{ (z-a_k)(z-b_b)} \,.
\ee
$g_e (z)$ and $g_o(z)$ are  holomorphic functions 
whose  large $z$ behavior is  $g_e (z) = O(z^0) $ and $g_o(z)= O(z^0)$. 
The loop equation \eqref{loop-G} is split  into two parts 
\begin{align}
& G_e(z)^2 +G_o(z)^2  = \varphi_2(z) \,,~~ G_e(z) G_o(z)   =  V_o'(z) H(z) \,.
\label{loop-split} 
\end{align} 
The way of splitting  is consistent with the fact that 
$H(z)$ has no cut along $\Lambda$. 

For  $n=2$, the non-trivial simplest case, 
the partition function satisfies the differential equations 
$ d_0 = v_0 ( -  \hbar^2\log Z_M ) $ and $ d_1 = v_1 ( - \hbar^2 \log Z_M ) $. 
It is convenient to re-parametrize as  $t= c_{3/2}/c_{1/2}^3$ and $x=c_{3/2}$
and regard the partition function  as the function of $t$ and $x$. 
The merit of parametrization is that any function of  $t$ is the homogeneous  
solution of  $v_0$ since $v_0 (t) =0$.  Employing the fact that
$v_0(x)= 3x/2$, $v_1(x)= 0$ and $v_1(t) =  -(3/2) t^2 c_{1/2}^2 $,
one has the differential equations
\be
d_0 = -  \frac  {3\hbar^2} {2}
x \frac{\partial} {\partial x }  ( \log Z_M )  \,,~~~~
\delta_1\equiv d_1/c_{1/2}^2  = \frac{3 \hbar^2}2  t^2  \frac{\partial} {\partial t }   (  \log Z_M )   \,.
\ee 
With  $d_0$ in \eqref{d_0}, we put $\log Z_M$ as
\be
\log Z_M = \frac {2 } {3\hbar^2}  \Big(   -  d_0\log x + Y(t)  \Big) 
\label{Y}
\ee
and $Y(t) $ satisfies the differential equation, 
$  t^2 {d Y(t)}/ {d t }    =\delta_1 $. 
This implies that  $ \delta_1 $ is the function of $t$ only
and vanishes as $t \to 0$. 

It is noted that $\varphi_2(z)$ in  \eqref{loop-split} is cast into the form 
$\varphi_2(z)=  {P_3(z)}/{z^5}  $ where
\begin{align}
P_3(z)=( d_0 +  \alpha^2) z^3 +  (d_1 + c_{1/2}^2) z^2 + 2 c_{3/2} c_{1/2}  z  +    c_{3/2}^2\,.
\label{loop-n=2} 
\end{align}  
Rescaling $z= u c_{1/2} ^2 $ and 
$ {P_3(z) } = c_{1/2}^6 Q_3(u)$, one puts 
$Q_3(u)$ in terms of $t$ and $x$, 
\begin{align}
Q_3(u) &=( d_0 +  \alpha^2) u^3 +  (\delta_1 +1 )u^2 + 2 t  u  +   t^2 
\nn\\
& \equiv ( d_0 +  \alpha^2) (u-A) (u-B) (u +\gamma^2 )  \,.
\label{Qu}
\end{align} 
$Q(u)=0$ has two positive roots $A$ and $B$ and one negative root 
 $-\gamma^2$. 
Two positive roots are related with branch points  
$a=A  c_{1/2} ^2$ and $b= B c_{1/2} ^2\,.$
The negative root is related with holomorphic function $g_e(z)$ and $ g_o(z)$
through the loop equation \eqref{loop-split}
\be
 g_e(z)^2 u  +{ g_o(z)^2}/{c_{1/2}^2} =  ( d_0 +  \alpha^2)( u + \gamma^2)\,.
\ee 
Noting that $ g_e(z)^2$ and $g_o(z)^2$ are even degree of $u$, 
we conclude that they are constant and 
$g_e(z)  = \sqrt{ d_0 +\alpha^2}$ and 
$ g_o(z)  =c_{1/2} { \gamma}  \sqrt{ d_0 +\alpha^2}$.

The relation between three roots are obtained from \eqref{Qu}
\begin{align}
& t^2  =( d_0 +  \alpha^2)\gamma ^2 AB\,,~~~
2 t = ( d_0 +  \alpha^2  )( AB - (A+B) \gamma^2 )\nn\\
& \delta_1  +1   =(d_0+\alpha^2 ) (\gamma^2   -  (A+B) ) 
\label{roots}\,.
\end{align}
In addition, $ H(z)$ in  \eqref{loop-split}  is the order of $1/z$
whose coefficient provides an additional identity;
$  \hbar b N + \alpha =  \gamma  (d_0+\alpha^2) $  
or $ \gamma  (\hbar bN+ \alpha)=1$.
This together with \eqref{roots} solves the algebraic equations
\begin{align}
AB&=t^2\,,~~  A+B= -2 t + t^2  (d_0 + \alpha^2)\nn\\
\delta_1   & =  2  (d_0+\alpha^2 ) t - (d_0+\alpha^2 )^2 t^2\,.
\end{align}
Therefore, $Y(t)$ in \eqref{Y}  is determined to the lowest order in $\hbar$  to give 
\begin{align}
&  \log( Z_M/\zeta )= \frac {2 } {3\hbar^2} 
 \Big(    -  d_0\log x   +2 (d_0 + \alpha^2 ) \log t 
- (d_0 + \alpha^2 )^2 t   \Big) 
\nn\\
& Z_M/ \zeta =  (c_{3/2})^ { 2( (bN + \hat \alpha)^2 + \hat \alpha^2) /3} 
\left(c_{1/2} \right)^ {- 4(bN + \hat \alpha)^2}
 \exp \left(  - \frac {2  c_{3/2} (bN+ \hat\alpha^2 )^4 \hbar^2  }  {3  c_{1/2}^3 }
\right)
\end{align} 
where $\hat \alpha = \alpha/\hbar$.

\section {Discussion and comments}
\label{discussion} 

We presented how to find the partition function of matrix model 
which corresponds to the irregular conformal block 
in the presence of the irregular singularity of odd degree.  
Explicit  solution is obtained for $n=1$ and  $n=2$ cases.   

It is obvious to generalize to  $n>2$. 
One may have 
$\varphi_2(z) = P_{2n-1}(z)/z^{2n+1}$ in  \eqref{loop-split}
where $P_{2n-1}(z)$ is  a polynomial of degree $(2n-1)$.
The $2(n-1)$  positive zeros of $P_{2n-1}(z) $ will provide the $2(n-1)$ 
branch points. The remaining factor linear in $z$ 
will  fix the holomorphic functions $g_e(z)$ and $g_o(z)$.  
$d_0$ is fixed as in \eqref{d_0} and other 
$d_k$'s are determined from the filling fractions, 
$ \oint_{\Lambda_k}  \frac{dz}{ \pi i}~ W(z) =\hbar b N_k $ with  $\sum N_k =N$.
It is noted that  the contour surrounding all $\Lambda_k$'s  is trivial.  
Instead, one additional constraint comes from the $1/ z$ behavior
 of  $ H(z)$ in \eqref{loop-split}.  

One may also consider more complicated potential 
which may contain polynomials together with inverse powers 
and a logarithmic one. Its partition function can be 
regarded as the inner product of the irregular states.  
Details of this consideration and the calculation of higher orders in $\hbar$ 
will appear elsewhere. 

Finally, it is noted that the matrix model we presented in this letter  reduces to the 
$O(n)$ matrix model on the random surface  
with $n=-2$ when $\beta=1$ \cite{eynard:1992}.   
It will be interesting to understand the results in terms of $O(n)$ matrix model. 

\subsection*{Acknowledgments} 
The author thanks Ivan Kostov and Takahiro Nishinaka for useful discussion.
This work is  supported in part by the National Research Foundation (NRF) of Korea funded by the Korea government (MEST) 2005-0049409.

\appendix 
\section{Explicit form of $f(z)$ } 
\label{f(z)}

We provide the explicit form of $f(z)$ presented in \eqref{f(z)-form}. 
Using the potential \eqref{Vz} one has 
\begin{align}
 f(z)& = 4g\sqrt{\beta} \sum_{I=1}^N\left\langle \frac{V'(z)-V'(\lambda_I)}{z-\lambda_I}\right\rangle
\nn\\
& =  4g\sqrt{\beta} \sum_{I=1}^N
\left\langle
\frac{\alpha (1/z- 1/\lambda_I ) + \sum_{s\in S} {c_s} (1/z^{s+1} -1/\lambda_I^{s+1})  } {z-\lambda_I}
\right\rangle 
\end{align}
where $S=\{1/2,3/2, \cdots, n-1/2\}$. 
Let us eliminate the $\alpha$ term by using the identity
\be
0= \left\langle  \sum_{I=1}^NV'(\lambda_I ) \right\rangle 
=  \sum_{I=1}^N \left\langle  \frac \alpha {\lambda_I} + \sum_s \frac{ c_s} { \lambda_I^{s+1}}  \right\rangle 
\ee
which stands for the invariance of the shift of the integration.  
One may wonder if the identity may fail since in our case, 
the integration is along the positive real line. 
However, this is not the case thanks to  the Vandermonde determinant 
because  the infinitesimal shift $\epsilon$ around the origin 
contributes to the order of $\epsilon^{N(N-1)/2}$ which does not contribute 
as $N \gg1$.  
\begin{align}
 f(z)& =-  4g\sqrt{\beta} \sum_{I, s} 
\left\langle
~\frac{c_s ( z^{s} -\lambda_I^{s})  } {z^{s+1} \lambda_I^{s} (z-\lambda_I)} 
\right\rangle
\nn\\
&= -   \frac{4g\sqrt{\beta} }{z^{3/2}} \sum_{I, s}
\left\langle
  \frac{c_s}{\lambda_I^s  }   \left(
\sum_{m =even \ge 0}^{2s-1} 
+\sum_{m =odd \ge  1}^{2s-2}   \right)   \frac {(\lambda_I /z )^{m/2}}{(\sqrt{z} + \sqrt{\lambda_I}) } 
\right\rangle \,.
\end{align} 
Inside the sum in even $m$,  one may add and subtract  $ 1/\sqrt{z} $ 
to $1/(\sqrt{z} + \sqrt{\lambda_I})$ to rewrite  
\begin{align}
 f(z)& = -  \frac{4g\sqrt{\beta}} {z^{2}}    \sum_{I, s} 
 \left\langle \sum_{m =even} \frac {c_s}{\lambda_I^{s-m/2} z^{m/2} } \right \rangle
\nn\\
&~~~ +  \frac{4g\sqrt{\beta} }  {z^{3/2}}   \sum_{I, s}
\left \langle  \frac {c_s }{ \lambda_I^{s} (\sqrt{z} + \sqrt{\lambda_I})}    
\left( \sum_{m =even}  \frac{\lambda_I ^{(m+1)/2}} {z^{(m+1)/2} } 
-  \sum_{m =odd} \frac{\lambda_I^{m/2}}{ z^{m/2} }  \right) 
\right\rangle 
\nn\\
&= -  \frac{4g\sqrt{\beta}} {z^{2}}    \sum_{I, s} 
 \left\langle\sum_{k=integer \ge0}^{s-1/2}  \frac1{\lambda_I^{s-k} z^{k} }   \right \rangle
+   \frac{4g\sqrt{\beta} }  {z^{3/2}}   \sum_{I, s}
\left \langle  \frac {c_s }{ z^{s} (\sqrt{z} + \sqrt{\lambda_I})}    
\right\rangle 
\end{align} 
Re-arranging the order of the summation over $s$ and $k$, one has 
\begin{align}
 f(z)&=  4g^2 \sum_{k=0} ^{n-1} \frac  { v_k (\log Z_M ) } {z^{2+k}}  
+ 4   g \sqrt{\beta} 
\left\langle  
\sum_I \frac1{ \sqrt{z} + \sqrt{\lambda_I} } 
\right\rangle ~ \frac{V_o'(z)}{\sqrt{z}}
\end{align}
where  $v_k = \sum_s  s c_{s +k}   \frac{\partial} {\partial c_s } $ 
with the definition $c_s =0$ if $s $ does not belong to $S$.


\begin{thebibliography}{9}

\bibitem {GT:2012} D. Gaiotto and J. Teschner, 
``Irregular singularities in Liouville theory and Argyres-Douglas type gauge theories, I", arXiv:1203.1052 [hep-th].

\bibitem {AGT:2009}  L. Alday, D. Gaiotto, Y.Tachikawa , 
``Liouville Correlation Functions from Four-dimensional Gauge Theories", 
 Lett. Math. Phys.91:167-197 (2010).

\bibitem{Bonelli:2011aa}
  G.~Bonelli, K.~Maruyoshi and A.~Tanzini,
  ``Wild Quiver Gauge Theories,''
  JHEP {\bf 1202} (2012) 031.

\bibitem{AD:1995} Philip C. Argyres and Michael R. Douglas, 
``New Phenomena in SU(3) Supersymmetric Gauge Theory",  
Nucl. Phys. {\bf  B448} (1995) 93. 

\bibitem{Gaiotto:0908} D. Gaiotto,
``Asymptotically free N=2 theories and irregular conformal blocks", 
arXiv:0908.0307  [hep-th]. 

\bibitem {TR:2012} T. Nishinaka and C. Rim, 
``Matrix models for irregular conformal blocks and Argyres-Douglas theories",
JHEP 10 (2012) 138.

\bibitem{Cecotti:2011rv}
  S.~Cecotti and C.~Vafa,
  ``Classification of complete N=2 supersymmetric theories in 4 dimensions,''
  arXiv:1103.5832 [hep-th].

\bibitem{Dijkgraaf:2009pc}
  R.~Dijkgraaf and C.~Vafa,
  ``Toda Theories, Matrix Models, Topological Strings, and N=2 Gauge Systems,''
  arXiv:0909.2453 [hep-th].

\bibitem{Eguchi:2009gf}
  T.~Eguchi and K.~Maruyoshi,
  ``Penner Type Matrix Model and Seiberg-Witten Theory,''
  JHEP {\bf 1002} (2010) 022.

\bibitem{Nishinaka:2011aa}
  T.~Nishinaka and C.~Rim,
  ``$\beta$-Deformed Matrix Model and Nekrasov Partition Function,''
  JHEP {\bf 1202} (2012) 114.

\bibitem{resolvent} 
See {\it e. g.\ } 
P. Ginsparg and G. Moore, " Lectures on 2D gravity and 2D string theory", arXiv:hep-th/9304011
and F. Di Francesco, P. Ginsparg and Z. Zinn-Justin,  arXiv:hep-th/9306153. 

\bibitem{eynard:1992} 
B. Eynard and J. Zinn-Justin, "The O(n) model on a random surface: critical points and large-order behaviour", Nucl. Phys. {\bf  B386} (1992) 558. 

\end{thebibliography}
\end{document}